# Wireless Networking to Support Data and Voice Communication Using Spread Spectrum Technology in The Physical Layer

Sourav Dhar[1] and Rabindranath Bera[2]

[1]Pailan College of Management & Technology, Pailan, Kolkata-700104
E-mail: dhar.sourav@gmail.com
[2]Sikkim Manipal Institute of Technology, Sikkim Manipal University, Majitar, Rangpo,
East Sikkim 737132, INDIA, E-mail: r_bera@hotmail.com

**Abstract**

*Wireless networking is rapidly growing and becomes an inexpensive technology which allows multiple users to simultaneously access the network and the internet while roaming about the campus. In the present work, the software development of a wireless LAN(WLAN) is highlighted. This WLAN utilizes direct sequence spread spectrum (DSSS) technology at 902MHz RF carrier frequency in its physical layer. Cost effective installation and antijaming property of spread spectrum technology are the major advantages of this work.*

## 1. Introduction

Mobile multimedia networking technology is growing at the lightning speed. People at different parts of the world are thinking of having wireless terminal to fulfill all their communication need. Particularly, the people at the hilly terrain little preference other than total wireless solution.

Thus the authors are encouraged to implement mobile virtual private network (MVPN) in an intelligent transport system (ITS). As a first step to this the laboratory work is done to establish a WLAN system which will be able to support the transmission and reception of voice, data, image and video.

Since security is the major concern hence spread spectrum technology has been used in the physical layer and MATLAB (version 6.5) has been used for software development.

In the laboratory, the data and voice transmission and reception have been successfully tested. Here the data has been assigned higher priority than voice, i.e., whenever data and voice transmission will be attempted simultaneously, the data will be transmitted suppressing voice. This leads to an "either or" situation [1]. Transmission and reception of both data and voice have been done suing the audio port of the computer. For the transmission and reception of data "chat window" is established using software.

The Telecom Regulatory Authority of India (TRAI) has set aside the 902 to 928 MHz, 2.4GHz and the 5.72GHz to 5.85 GHz bands for unlicensed transmissions [2]. The 902MHz range has been the most widely used because of its cost effectiveness [3]. Hence, in this work 902MHz RF carrier frequency is chosen.

The arrangement for wireless communication between two computers and the handshaking process have already been presented in [1].

Here authors would like to present the MATLAB (version 6.5) based software development and the relevant results.

## 2. Software development

For software development we are using MATLAB (version 6.5). Using this software we have established the "chat mode". So the arrangement supports more user friendly approach. So far all the characters have been coded. Coding is done using dual tone multiple frequency (DTMF) encoding scheme. The frequencies corresponding to each character has been shown in table-1. This approach supports us to use the audio ports for interfacing the telephone set with computers.

In the software part, a look up table is generated where the two frequencies corresponding to each character are listed with a tolerance of ±5%. The same program is used for transmission as well as for reception. The chat window is established using graphic user interface (GUI) of MATLAB (version 6.5). the chat window is shown in fig -1.

At the transmitter side, the character to be sent is typed in the transmit window. Comparing the character with the characters of the lookup table, corresponding frequencies are generated using software and then transmitted through the audio port of the computer. The arrangement for the communication between two computers is shown in fig-2. [1].

At the receiver side, after receiving the signal through the microphone input of the computers, its FFT is done. Peaks of the spectrum are detected and the frequencies of the peaks of the spectrum are estimated online. Then these frequencies are compared with the frequencies of the look up table and the corresponding characters are displayed in the receive window.

For establishing the look up table , the frequencies are not taken arbitrarily, instead these totally depend on the sampling frequency. Here the sampling frequency chosen is 8000Hz . to code the numbers and the characters , we have used lower range frequencies in the range of 699Hz to 990Hz and upper range frequencies in the range of 1151Hz to 1497Hz as shown in table-1. The frequencies for coding are found by means of a hit and trial method. These frequencies can be changed if a different sampling frequency is used.

## 2.1. Concept of codec program

For handshaking, the source will transmit a code consisting source & destination address, a start bit and an acknowledgement bit. This is a 14 bit code and generated by software program. The code format is given in fig-3.

First start bit is made 1 for enabling transmission. And the acknowledgement bit is 0 for the transmitter wish to communicate. On reception of this code the destination will interchange the source and destination addresses and will make acknowledgement bit 1, and then retransmit the code. Hence the connection will be established between two computers.

"1" is coded by a pulse having pulse repetition time (PRT) 800μs and "0" is coded by a pulse having PRT 600μs. these pulses are of 50% duty cycle. And pulse amplitude is made 20mv.

Now let us take an example to make it clear. Computer 8 wishes to communicate with computer 1. Then the source address will be 001000 and destination address will be 000001. The source address is stored in a register of the microcontroller. There is a lookup table where the addresses of the other computers are stored. Thus on reception of a start command the codec (microcontroller) of the transmitting computer8 generates a 14 bit code with acknowledgement bit 0.

On reception of this code, codec of computer 1 will first check the destination address, if it matches with its address that is already stored in its one of the registers, then it interchanges the source and destination addresses and makes acknowledgement bit 1 and then retransmits the code. This code is transmitted serially i.e. in a sequential fashion.

As this radio unit uses frequency shift keying (FSK) modulation, thus this code will be FSK modulated in the radio unit at 902MHz and then transmitted. At the output of the radio unit we will have compulsorily a sine wave. Same sinusoidal analog signal will be received by the radio unit of the computer 8. But the codec cannot deal with the sinusoidal signals. Thus to decode the received signal, it should be converted into square wave first to provide to discrete level. The simplest of the sine to square wave converters is a comparator used as a zero-crossing detector. Thus pulse recovery is done.

But pulse recovery is not code recovery. The code recovery is done by actual measurement of pulse repetition time (PRT). And a software based decision will be taken by the microcontroller.

This code recovery is done within the microcontroller. Hence this is entirely software based. There is a counter in the microcontroller which is enabled by the negative edge trigger. As soon as a negative edge comes in , the counter will start counting it will count till another negative edge comes in. the count value will be latched and stored into one of the registers. Time for one count will be decided by the frequency of the crystal that is used for providing clock to the microcontroller. After measurement of PRT , the result is given to the decision device. A threshold level is set there, in this case we set threshold at 700µs. if the PRT is less than 700µs then it is zero and if it is more than 700µs then it is 1. Thus the code is recovered.

### 3. Performance of the system

The system is working fine as the intended data is received through a remote mobile PC while the data is sent through a base station PC. This has been achieved by putting off the sweep RF source. As soon as the sweep RF source is turned on the intended data is lost for a period of time. The data lost period is highly dependable on the sweep time as well as on the power level of the sweep generator. As the sweep time varies the relative dwell time also varies; resulting different jamming occurrences of the radio channel at different power level. Table-2 shows some experimental data regarding dwell time vs. required jamming power whereas fig-4 shows its graphical representation. It is evident from the plot that as dwell time increases the required jamming power decays exponentially following some empirical formula like this – Fit equation is , $y0 + A_1 e^{-(x-x0)/t1} + A_2 e^{-(x-x0)/t2}$ where co-efficient are shown in the plot itself ( fig.-4) [4].

The high power jamming problem beyond DSSS jamming margin is solved by utilizing frequency diversion technique of the carrier frequency. Since frequency diversion technique is implemented here, on realizing the jamming condition the micro controller shifts the synthesized frequency generator to next available free radio channel. During this process some data packet may be lost which can recovered by some signal processing techniques like Automatic Repeat Request (ARQ ). The process of diversion continues to next available free channel as soon as that particular frequency comes out from the sweep generator.

### 4. Conclusion

It is a well established fact that spread spectrum based wireless technology is coming with great hope to support user need in the form of multimedia service compared to existing technology. It has started to support high quality voice in mobile environment without any multimedia service. We are trying to exploit this fact towards the multimedia communication. Due to its high anti jamming quality spread spectrum technology is essential in the mobile computing field for its reliable operation.

## 5. Acknowledgement

The authors are thankful to the technical assistants of E&C Engineering Department, SMIT, Prof. A.Choudhury, Dean (Academic), SMIT, Prof. A.C. Majumder, Principal Director, PCMT and Prof. Shouvic Roy, HOD, IT Deptt. ,PCMT for providing their support both technically and financially.

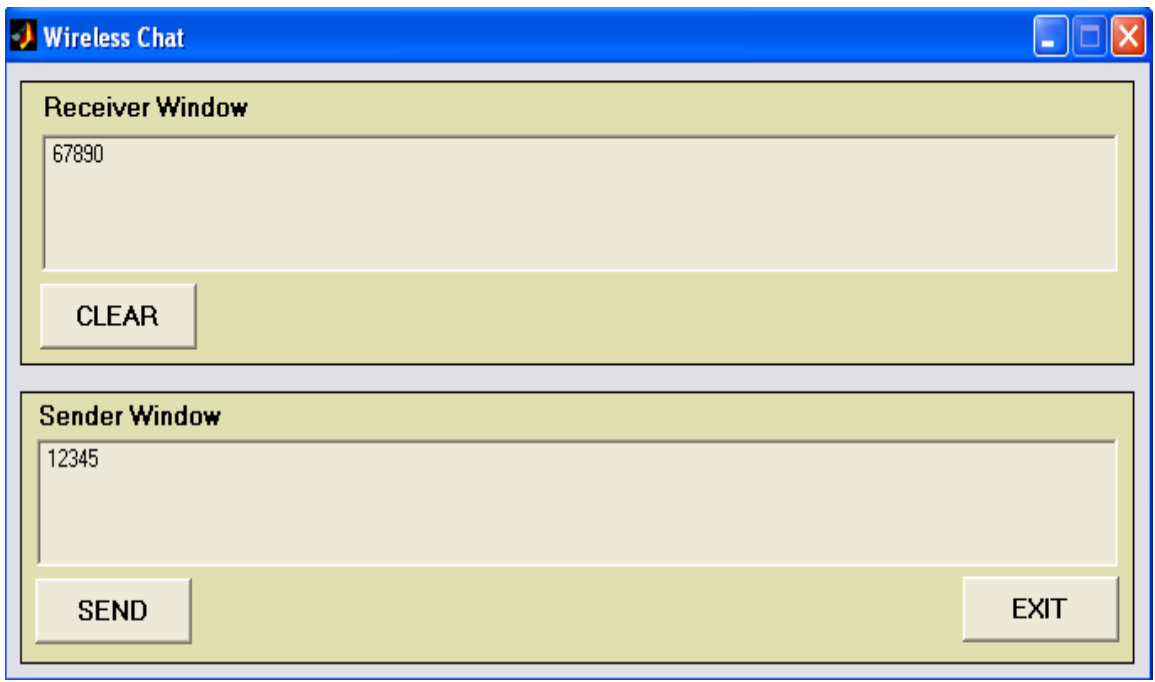

Fig -1: Chat window

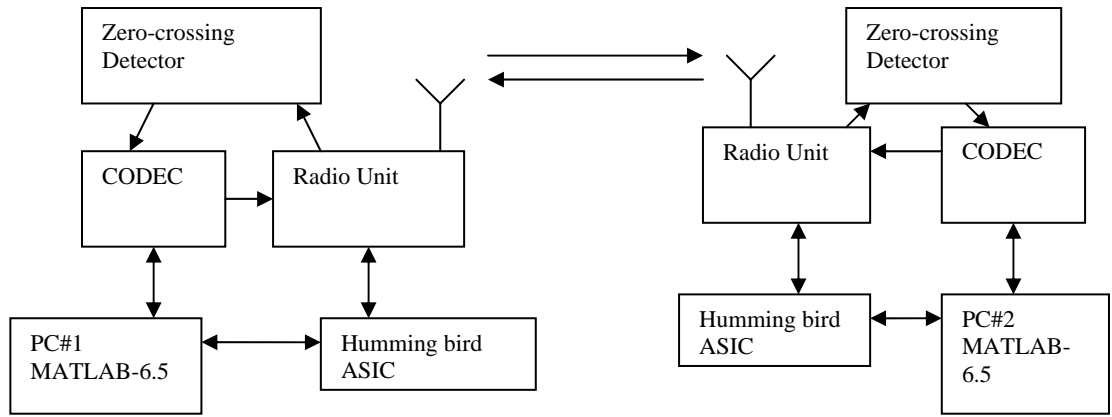

Fig-2: Arrangement for wireless communication between two computers.

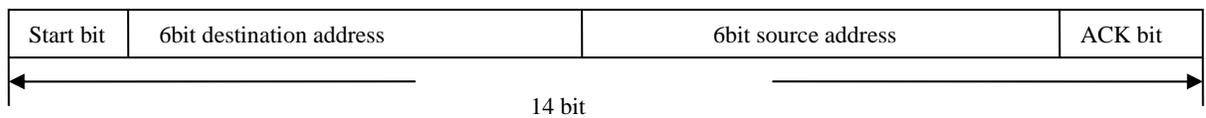

Fig-3: code format (14 bit)

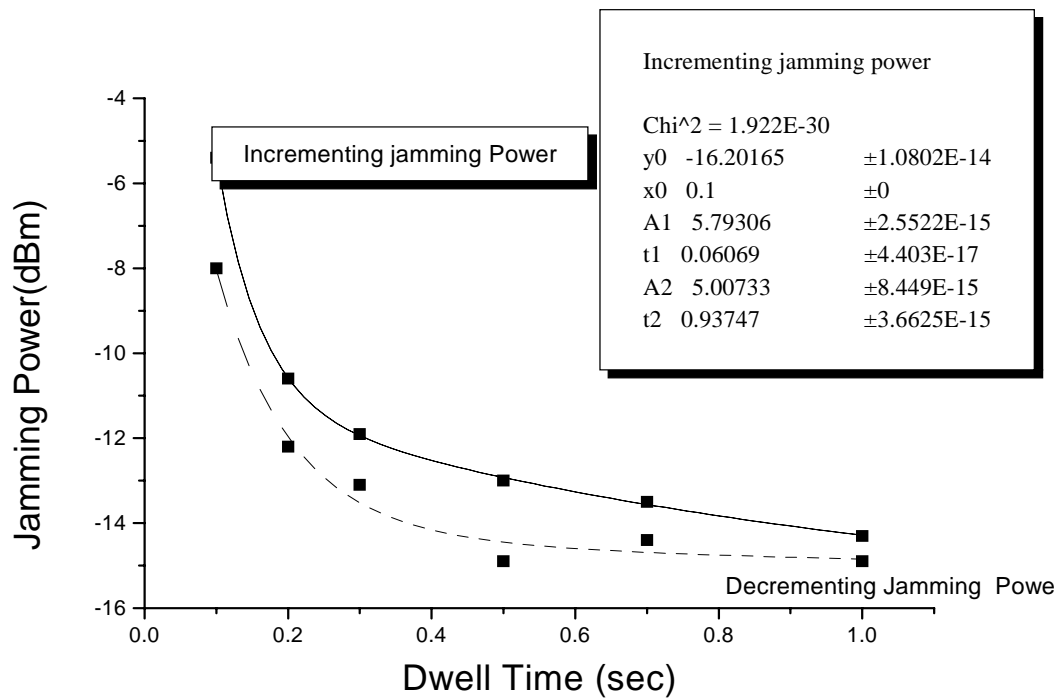

Fig-4: Dwell Time Vs. Jamming Power graph

| Frequency in Hz | 699 | 772 | 842 | 854 | 869 | 880 | 918 | 930 | 943 | 990 |
|---|---|---|---|---|---|---|---|---|---|---|
| 1151 | **0** | **1** | **2** | **3** | **4** | **5** | **6** | **7** | **8** | **9** |
| 1168 | **A** | **B** | **C** | **D** | **E** | **F** | **G** | **H** | **I** | **J** |
| 1179 | **K** | **L** | **M** | **N** | **O** | **P** | **Q** | **R** | **S** | **T** |
| 1211 | **U** | **V** | **W** | **X** | **Y** | **Z** | **a** | **b** | **c** | **d** |
| 1236 | **e** | **f** | **g** | **h** | **i** | **j** | **k** | **l** | **m** | **n** |
| 1280 | **o** | **p** | **q** | **r** | **s** | **t** | **u** | **v** | **w** | **x** |
| 1384 | **y** | **z** | **:** | **;** | **[** | **]** | **{** | **}** | **"** | **<** |
| 1369 | **>** | **?** | **,** | **.** | **/** | **+** | **-** | **\*** | **\\** | **\|** |
| 1451 | **~** | **!** | **@** | **#** | **$** | **%** | **^** | **&** | **_** | **(** |
| 1497 | **)** | **=** |   |   |   |   |   |   |   |   |

Table-1: look up table for DTMF encoding.

| Dwell Time (sec) | Increasing Jamming Power (dBm) | Decreasing Jamming Power (dBm) |
|---|---|---|
| 0.1 | -5.4 | -8 |
| 0.2 | -10.6 | -12.2 |
| 0.3 | -11.9 | -13.1 |
| 0.5 | -13.0 | -14.9 |
| 0.7 | -13.5 | -14.4 |
| 1.0 | -14.3 | -14.9 |

Table-2: Experimental data – Dwell time vs. jamming Power